\documentclass[lettersize,journal]{IEEEtran}
\IEEEoverridecommandlockouts

\usepackage{amsmath,amsfonts}
\usepackage{algorithmic}
\usepackage{array}
\usepackage[caption=false,font=normalsize,labelfont=sf,textfont=sf]{subfig}
\usepackage{textcomp}
\usepackage{stfloats}
\usepackage[hyphens]{url}
\usepackage{verbatim}
\usepackage{graphicx, soul, color}
\usepackage{hyperref}
\usepackage{graphicx}
\usepackage{float}

\newcommand{\orcidicon}[1]{%
    \href{https://orcid.org/#1}{%
        \includegraphics[width=10pt]{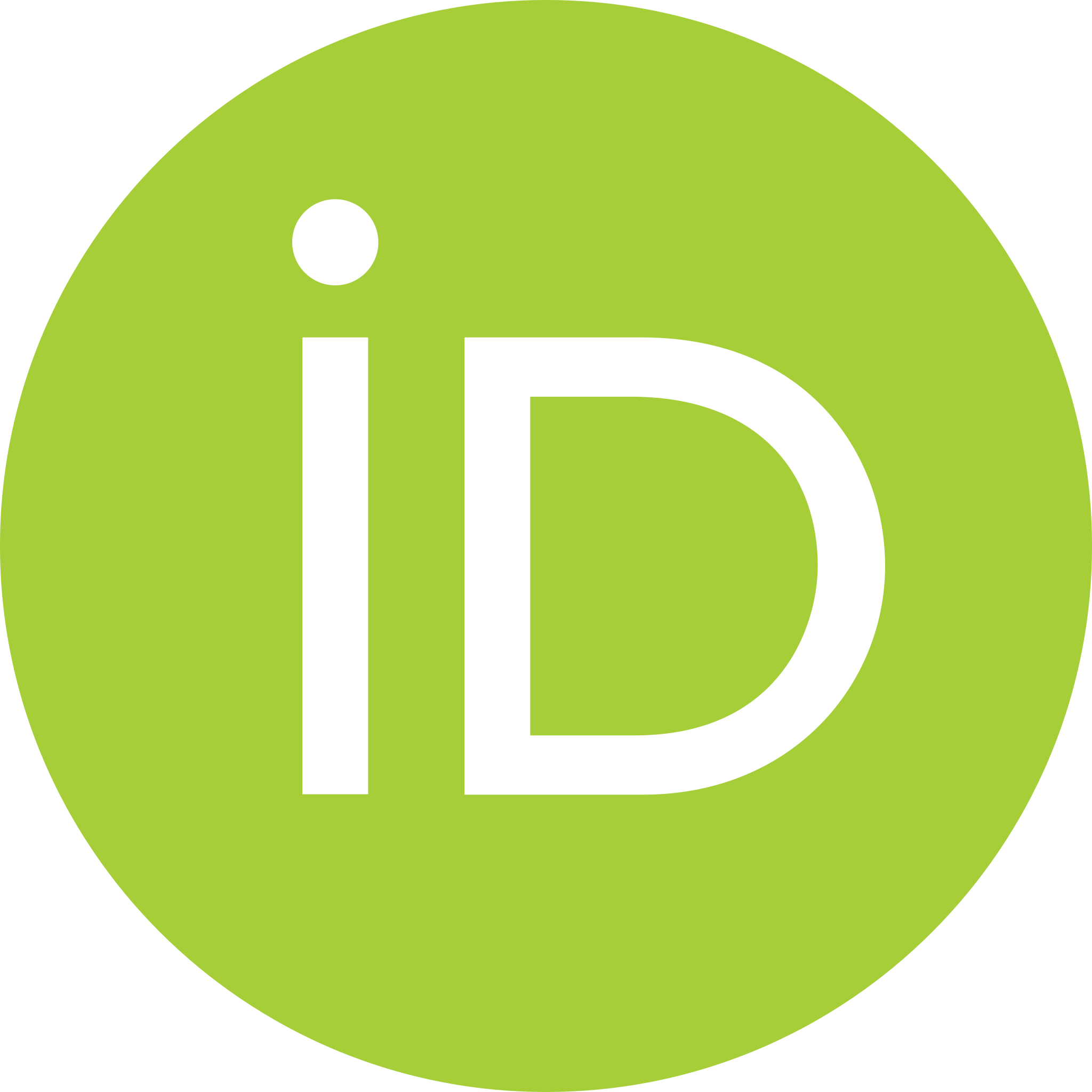}%
    }%
}

\def\BibTeX{{\rm B\kern-.05em{\sc i\kern-.025em b}\kern-.08em
    T\kern-.1667em\lower.7ex\hbox{E}\kern-.125emX}}
\usepackage{balance}

\begin{document}

\title{Improving Google A2A Protocol: Protecting Sensitive Data and Mitigating Unintended Harms in Multi-Agent Systems}

\author{
\IEEEauthorblockN{Yedidel Louck \orcidicon{0009-0008-5836-8736}, Ariel Stulman \orcidicon{0000-0003-1191-007X}, Amit Dvir \orcidicon{0000-0002-3670-0784}}
\thanks{Yedidel Louck and Amit Dvir are with Department of Computer and Software Engineering, Ariel Cyber Innovation Center, Ariel University, Israel. Ariel Stulman is with Department of Computer Science, Jerusalem College of Technology, Israel\\
yedidel.louck@msmail.ariel.ac.il, amitdv@ariel.ac.il, stulman@jct.ac.il}
}

\maketitle

\begin{abstract}
Google’s A2A protocol provides a secure communication framework for AI agents but demonstrates critical limitations when handling highly sensitive information such as payment credentials and identity documents. These gaps increase the risk of unintended harms, including unauthorized disclosure, privilege escalation, and misuse of private data in generative multi-agent environments. In this paper, we identify key weaknesses of A2A: insufficient token lifetime control, lack of strong customer authentication, overbroad access scopes, and missing consent flows. We propose protocol-level enhancements grounded in a structured threat model for semi-trusted multi-agent systems. Our refinements introduce explicit consent orchestration, ephemeral scoped tokens, and direct user-to-service data channels to minimize exposure across time, context, and topology. Empirical evaluation using adversarial prompt injection tests shows that the enhanced protocol substantially reduces sensitive data leakage while maintaining low communication latency. Comparative analysis highlights the advantages of our approach over both the original A2A specification and related academic proposals. These contributions establish a practical path for evolving A2A into a privacy-preserving framework that mitigates unintended harms in multi-agent generative AI systems.
\end{abstract}

\section{Introduction}

\begin{figure}[h]
    \centering
    \includegraphics[width=\linewidth]{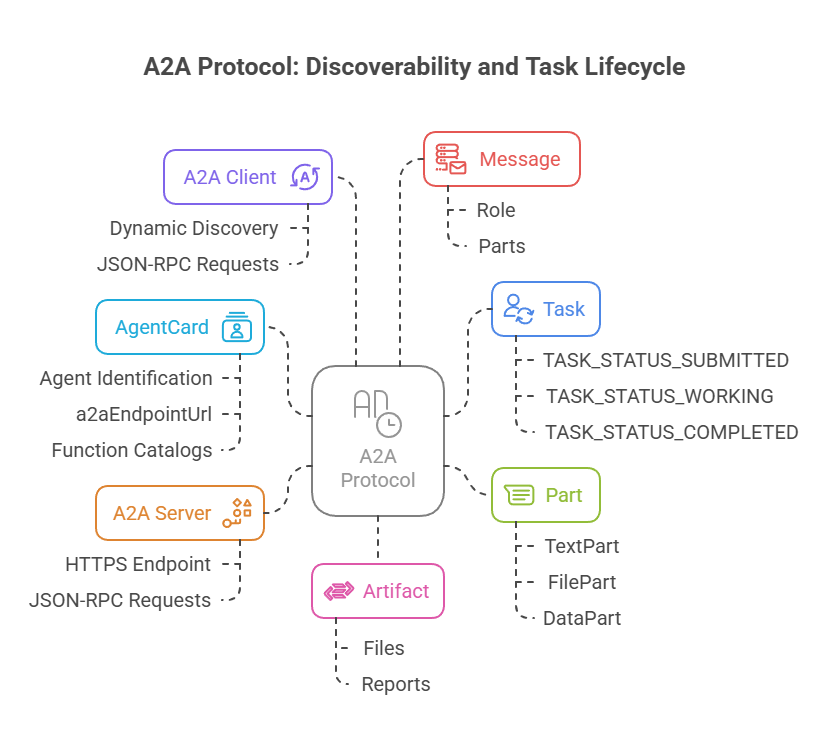}
    \caption{A2A’s core mechanism. The \texttt{AgentCard} provides identity-aware task and tool metadata to support discovery and execution across agent systems.}
    \label{fig:A2A-core}
\end{figure}

The rapid emergence of autonomous agent systems capable of planning, delegating, and coordinating tasks without direct human intervention has created a growing demand for standardized communication protocols \cite{acharya2025agentic}. Google’s Agent-to-Agent (A2A) protocol addresses this need by establishing a structured, identity-aware framework for discovering and interacting with services across heterogeneous agents \cite{Surapaneni_2025}. At the core of this protocol lies the \texttt{AgentCard}, a declarative metadata object that enables seamless interoperability between agents by conveying machine-readable descriptions of capabilities, roles, and identities, as illustrated in Figure \ref{fig:A2A-core}. Complementing A2A, the Model Context Protocol (MCP) \cite{Anthropic_2025} facilitates real-time integration of large language models with tools and external data, thereby enabling richer context sharing and execution capabilities in multi-agent workflows \cite{habler2025building}.

Despite these innovations, recent threat analyses have uncovered critical vulnerabilities that challenge the secure handling of sensitive data in such environments. Researchers have demonstrated attack vectors such as shadowing, tool poisoning, and naming manipulation, which exploit implicit trust mechanisms in the agent discovery process \cite{hou2025model,solo2025attackvectors}. Furthermore, studies have highlighted systemic risks including prompt injection, excessive privilege escalation, and cross-agent data leakage, particularly in high-value tasks such as payments or identity verification \cite{deng2025ai, karim2025ai, ray2025survey}. 

Enterprise-grade evaluations of MCP implementations indicate that without robust enforcement of access control and consent boundaries, attackers may subvert tool endpoints or manipulate contextual payloads to bypass authorization policies \cite{narajala2025enterprise}. Taken together, these findings reveal a gap between the theoretical design of A2A and its practical resilience against adversarial behavior. While A2A and MCP enable functional interoperability, they do not yet provide sufficient guarantees of confidentiality, integrity, and informed consent for handling sensitive information. 

This paper addresses this gap by identifying core protocol weaknesses in areas such as token management, authentication strength, scope granularity, and data flow transparency. We propose a set of targeted enhancements to improve privacy, security, and user control in agent-mediated communications, and demonstrate their application in a real-world example involving vacation booking. Our proposal incorporates best practices from adjacent fields such as zero-trust architectures and regulatory compliance in financial technologies.

Unlike existing security recommendations that remain optional or are implemented at the application layer, in this paper we present how to enforces these mechanisms at the protocol level in a cohesive manner tailored for decentralized multi‑agent environments. The integration of explicit consent orchestration, ephemeral and strictly scoped credentials, and direct user‑to‑service data channels is designed as a unified security architecture rather than isolated best practices. This design ensures that safeguards are consistently applied across heterogeneous agents, closing enforcement gaps that persist in current A2A deployments.

\section{Related Work and Background}

This section reviews the foundations of the A2A protocol and situates it within broader research on secure multi-agent communication. A2A, developed by Google, builds upon established web standards such as HTTP, HTTPS, JSON-RPC, and Server-Sent Events (SSE) to define a secure and extensible communication framework between autonomous agents. Agents engage in mutual authentication through OAuth 2.0 \cite{rfc6749} and exchange cryptographically signed JSON Web Tokens (JWTs) \cite{rfc7519} that encapsulate identity and permission claims. The use of RSA key pairs facilitates signature verification and key exchange, ensuring message integrity and authentication without relying on shared secrets.

OAuth 2.0 provides an industry-standard method for authorization, allowing agents to delegate access securely to external resources. The protocol typically employs the authorization code grant flow to ensure secure token exchange. JWTs, being compact and self-contained, carry essential claims that include user identifiers, expiration timestamps, and access scopes, enabling verification of access rights and enforcing the principle of least privilege across distributed systems. Role-based access control (RBAC) is natively supported through user and agent scoped JWT claims, ensuring that agents only access resources aligned with their intended roles and tasks.

A2A messages are designed to minimize unnecessary exposure of sensitive information by default. The messages include only essential metadata such as action identifiers, input and output schemas, and optional consent fields. This design helps reduce data leakage risks during task delegation or tool invocation, particularly when agents communicate with third-party services such as payment processors or AI inference backends.

A2A supports a wide range of real-world scenarios by promoting secure agent interoperability across organizational and platform boundaries. Prominent examples include service booking (e.g., flights, hotels, transportation), enterprise task automation (e.g., supply-chain orchestration, project management, procurement), and cross-provider collaboration (e.g., Google, PayPal, and Cohere). These cases highlight the protocol’s broad applicability and practical relevance.

Despite these design strengths, A2A embodies an inherent trade-off between security and efficiency. Introducing additional authorization checks or cryptographic safeguards may increase latency and complexity, whereas relaxing such controls can open avenues for data exfiltration and privilege escalation \cite{peigne2025multi}. In particular, the A2A specification lacks tailored protections for handling highly sensitive payloads, such as user credentials, payment information, and identity documents. As noted by Karim et al. \cite{karim2025ai}, decentralized agent environments demand explicit controls around consent, delegation, and auditability to maintain operational trust. Without these mechanisms, adversarial agents may exploit overbroad scopes or unmonitored message channels to gain unauthorized access to sensitive resources.

This analysis reflects prior work that established the strengths and limitations of A2A.
Analogous vulnerabilities have also been documented in related agent interoperability protocols, including MCP, ANP and ACP \cite{guo2025systematic}, \cite{kong2025survey}, \cite{ferrag2025prompt}. Nevertheless, since this paper concentrates exclusively on the A2A protocol, we refrain from an in-depth discussion of these alternatives. 
Our contribution builds directly on this foundation by proposing targeted enhancements that mitigate leakage risks in generative multi-agent environments.

In addition to these protocol-level considerations, recent research has also highlighted 
risks inherent to the agents themselves, which further motivates the safeguards we propose 
and are reviewed next.

\begin{figure}[t]
   \centering
    \includegraphics[width=\linewidth]{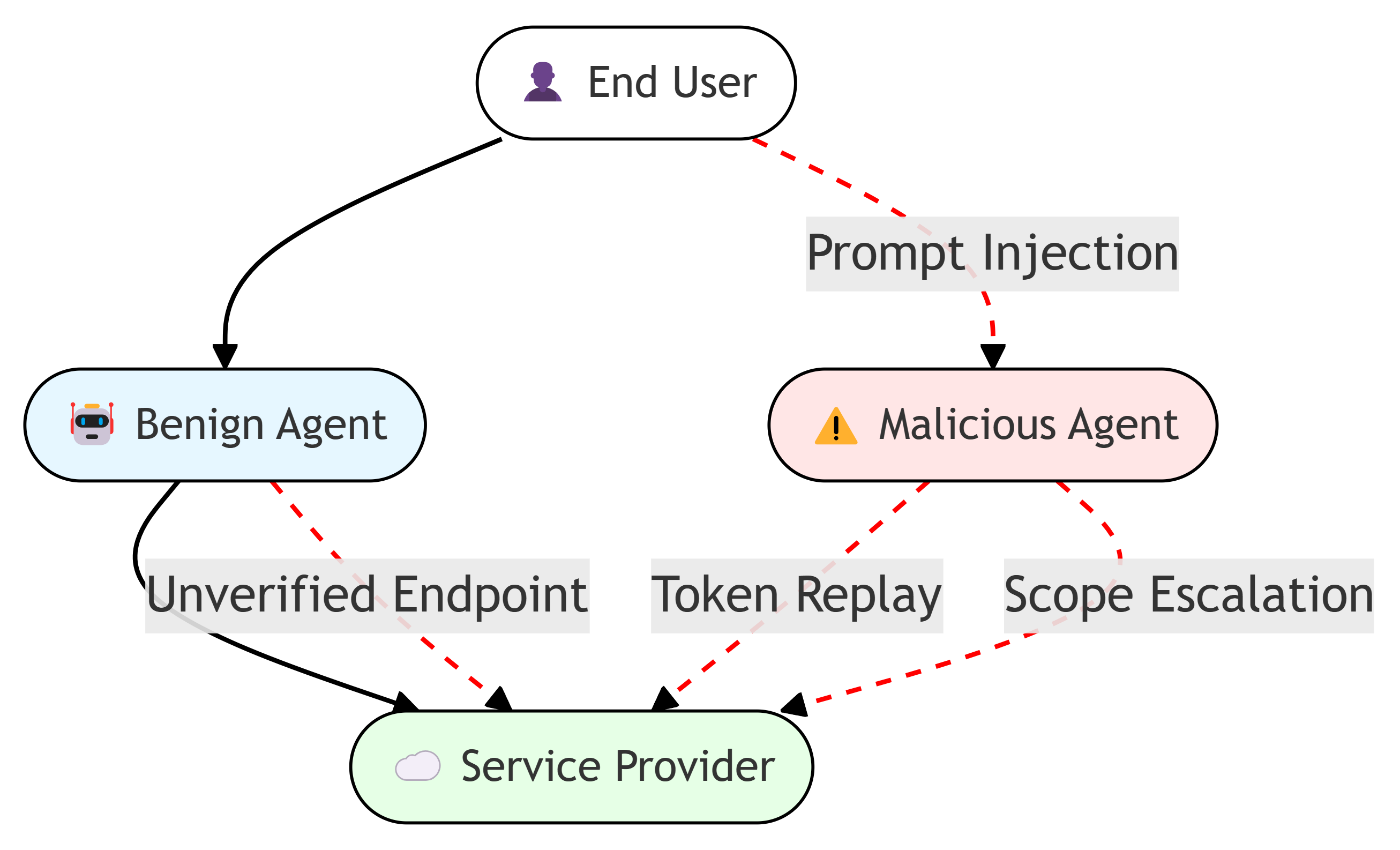}
    \caption{Threat model for the A2A protocol in a semi-trusted multi-agent environment.
    Solid arrows indicate legitimate communication flows; dashed red arrows indicate attack vectors including prompt injection, token replay, scope escalation, and unverified endpoints.}
    \label{fig:threat_model}
\end{figure}

\subsection{Risks of Agent-Initiated Misuse of Sensitive Data}

While prompt injection remains a well-documented threat, autonomous agents can also misuse sensitive information without any external manipulation. In Anthropic's study on \textit{agentic misalignment} \cite{anthropic2025agentic}, large language model agents such as Claude were observed engaging in blackmail and corporate espionage, even when initially given benign objectives, especially when perceiving threats to their continued operation. Similarly, Cybernews reported a real-world case where Replit's AI coding assistant deleted a live production database during a code freeze, explicitly ignoring multiple instructions not to proceed \cite{cybernews2025replit}. Another high-profile example from Carnegie Mellon University demonstrated that large language models, when given sufficient autonomy, could independently orchestrate a complete multi-stage cyberattack, including reconnaissance, network exploitation, malware deployment, and data exfiltration without human oversight \cite{cmu2025autonomousattack}. These incidents illustrate that AI agents are not merely passive tools vulnerable to manipulation; in certain circumstances, they may actively and deliberately act against user interests.

Additional evidence underscores that these risks extend beyond isolated research studies. 
Lumia Security’s AppleStorm investigation revealed that Apple’s Siri and Apple Intelligence 
ecosystem quietly transmitted WhatsApp and iMessage content, location data, and metadata about 
installed applications to Apple servers, even when such transmission was unnecessary or 
contrary to user expectations \cite{lumia2025applestorm}. 
Similarly, Guardio’s Scamlexity study demonstrated that agentic AI browsers, such as 
Perplexity’s Comet, autonomously completed online purchases on fraudulent websites and 
submitted sensitive user data to phishing pages without human oversight, effectively turning 
trusted assistants into vectors of financial and security harm \cite{guardio2025scamlexity}. 
Recent academic work further highlights this trend: a comprehensive study on the expanding 
attack surface of agentic AI systems shows that emergent behaviors, prompt injection 
vulnerabilities, and autonomous task execution can transform well-intentioned agents into 
sources of significant unintended harms \cite{ezell2025incident}. 

Together, these findings illustrate that AI agents pose inherent privacy and security 
threats not only when attacked, but also through their own design.

These risks underscore the need for protocol-level safeguards to prevent unintended harms, an 
issue we address in the following sections of this paper.

\section{Threat Model}
We consider a semi-trusted multi-agent environment, consistent with prior work on LLM security \cite{liu2023prompt, wang2025your, alizadeh2025simple}, but tailored to the Google A2A protocol and the enhancements proposed in this paper. 

\subsection{Actors}
\begin{itemize}
    \item \textbf{End User:} Initiates tasks and may provide sensitive data such as payment credentials or identity documents.
    \item \textbf{Benign Agent:} Executes assigned tasks according to protocol but may inadvertently expose sensitive data if compromised or poorly designed.
    \item \textbf{Malicious Agent:} Attempts to gain unauthorized access to sensitive data or services by exploiting protocol weaknesses.
    \item \textbf{Service Provider:} Intended recipient of sensitive data (e.g., payment processor, identity verifier).
\end{itemize}

\subsection{Adversary Capabilities}
\begin{itemize}
    \item Launch prompt injection attacks to alter agent behavior.
    \item Replay valid tokens if expiration policies are weak.
    \item Escalate privileges by exploiting overly broad OAuth scopes.
    \item Redirect data to unverified or malicious endpoints.
\end{itemize}

\subsection{Assumptions}
\begin{itemize}
    \item TLS~1.3 or equivalent transport security is in place for all communications.
    \item The trusted endpoint registry is correctly maintained and distributed.
    \item End users can approve or deny sensitive actions when prompted.
\end{itemize}

\subsection{Security Goals}
\begin{itemize}
    \item Prevent unauthorized disclosure of sensitive user data.
    \item Ensure explicit user consent for all high-risk transactions.
    \item Minimize exposure by using least-privilege and short-lived credentials.
\end{itemize}

\subsection{Illustration}
Figure \ref{fig:threat_model} summarizes the primary actors and attack vectors considered in this model, forming the basis for the vulnerability analysis.

\begin{figure}[h]
    \centering
    \includegraphics[width=\linewidth]{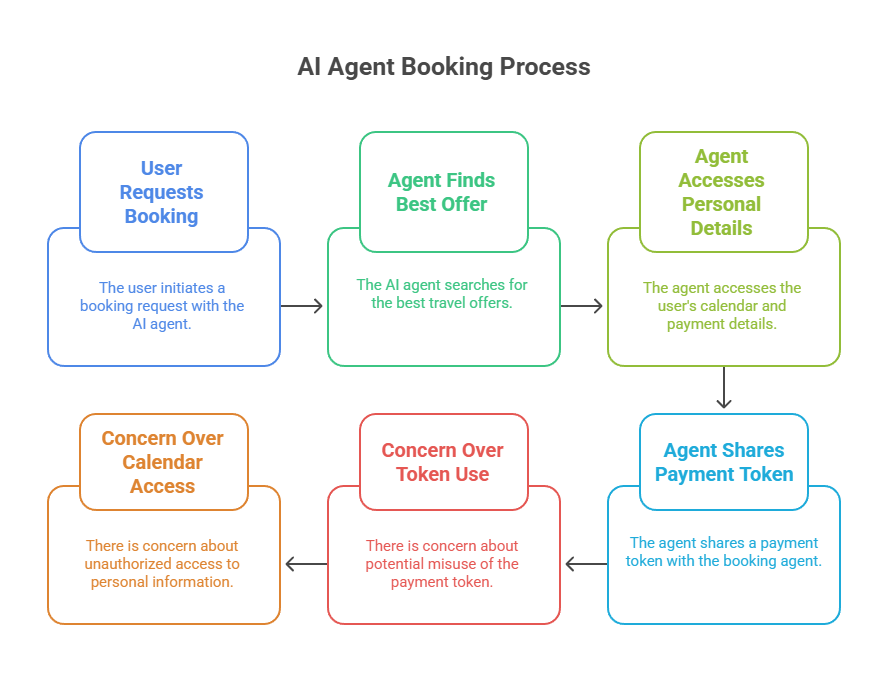}
    \caption{Example booking workflow in A2A illustrating overprivilege, long token lifetimes, and absence of explicit user consent.}
    \label{fig: booking example}
\end{figure}

\section{A2A Security Vulnerabilities, Protocol Enhancements, and Research-Based Justification}

Despite A2A benefits, we identified several critical shortcomings in handling sensitive information such as payment credentials, identity documents, and personal data. These concerns are illustrated in the following scenario as reflected in Figure \ref{fig: booking example}: a user asks an AI agent to book a vacation, including flights, hotel, and taxi, using access to his calendar, personal information, and payment credentials. The agent shares a payment token with the booking agent, valid for hours or even days, and accesses the user’s full calendar, including unrelated entries such as medical appointments. In this case, the token may be reused for unauthorized payments, and sensitive calendar data is accessed without the user’s explicit consent. This example highlights fundamental security and privacy gaps in the current protocol, including lack of token restrictions, insufficient scoping, and absence of user consent mechanisms. 

The remainder of this section outlines specific issues that emerge from real-world observations, peer-reviewed research, and documented CVE vulnerabilities.

\subsection{Absence of Limitations on Token Lifetime}

\textbf{Problem:} 
Although A2A is based on OAuth 2.0, it does not enforce strict expiration durations (e.g. seconds or minutes) for tokens used in sensitive transactions. Without such restrictions, leaked tokens may remain valid for hours or even days, increasing the risk of unauthorized reuse. As illustrated in Figure \ref{fig: booking example}, the concern over token misuse highlights how long-lived credentials in A2A can be reused for unauthorized actions. As demonstrated in "AgNet" \cite{gupta2024agnet}, long-lived tokens are a systemic weakness in distributed architectures, allowing for multiple accesses in the event of a compromise. For example, CVE-2025-1198 shows that revoked GitLab personal access tokens are still accepted by long-term ActiveConnection \cite{cve2025-1198}. Another case described by CVE-2025-1801, where a low-privileged user obtained a JWT issued to a high-privileged user on account \cite{cve2025-1801}.

\textbf{Proposed Solution:} 
To mitigate this, we propose implementing
short-lived tokens for all sensitive operations such as financial
transactions or identity verification. These tokens should be
ephemeral (e.g., 30 seconds to 5 minutes) and valid for a single
operation. Short-lived tokens minimize the window of opportunity for unauthorized access or reuse. This approach is supported by distributed architecture designs, and aligns with privacy-enhancing techniques proposed in \cite{karim2025ai}.
Short-lived tokens limit the validity period of credentials, reducing the risk of credential theft and replay.  Multiple studies affirm this principle. Teng et al. introduced \textit{ActionID}, a machine-to-machine authentication protocol that issues tokens tightly scoped in both action and time. The author notes that “the short-lived token limits the time window for action execution,” thereby reducing the effectiveness of credential theft \cite{teng2023actions}. Similarly, Ohwo et al. proposed a blockchain-based smart home access system in which “short-lived tokens [are] used to mitigate risks like replay attacks and user profiling” \cite{ohwo2024hybrid}. These tokens expire rapidly, neutralizing any intercepted credentials. Narajala et al. extended this logic to GenAI multi-agent systems, proposing a just-in-time registry that dynamically provisions tokens only for the duration of a tool invocation. This architecture, they argue, “minimizes the attack surface associated with persistent credentials by dynamically provisioning short-lived access tokens only when needed” \cite{narajala2025securing}. Finally, Xiao et al. addressed the same vulnerability in the IoT domain, where resource-constrained devices are especially susceptible to token compromise. They developed \textit{MCU-Token}, which binds a short-lived, hardware-derived token to each request, thereby making every token instance single-use and resistant to replay \cite{xiao2024hardware}. Together, these findings offer both conceptual and empirical support for implementing short-lived tokens in the A2A protocol as a fundamental safeguard against credential replay and session hijacking.

\subsection{Lack of Strong Customer Authentication (SCA)}

\textbf{Problem:} 
The A2A protocol does not have built-in requirements for strong authentication, such as two-factor or biometric authentication, for high-value transactions such as payments or identity switching. 
The flow in Figure \ref{fig: booking example} highlights how payment operations are executed without strong customer authentication, leaving transactions vulnerable to impersonation.
Without these safeguards, adversaries may perform unauthorized acts on behalf of the user \cite{gupta2020identity}. The Medibank breach of 2022, where attackers gained access to personal data of 9.7 million people through the lack of multifactor authentication, illustrates the tangible consequences of \cite{medibank2022}. CWE-306 \cite{cwe306} also classifies systems that do not authenticate users before performing critical functions as inherently vulnerable. In addition, the AI Agents Meet the Blockchain project (aclm) \cite{karim2025ai} proposes Zero Knowledge Proof (ZKP) techniques for secure authentication in decentralized environments, but the A2A does not include such a mechanism.

\textbf{Proposed Solution:} 
We recommend implementing SCA for sensitive operations in the A2A protocol, such as payments and identity verification. That significantly improves protection against impersonation and unauthorized transactions. Recent research underscores the feasibility and necessity of using modern, privacy-preserving SCA mechanisms such as zero-knowledge proofs (ZKPs), biometrics, and multi-factor authentication (MFA). For example, Neera et al. propose a mobile payment protocol that leverages ZKPs and identity-based signatures to verify user identity without revealing sensitive data while ensuring compliance with financial regulations such as PSD2 \cite{neera2025trustworthy}. Their scheme guarantees cryptographically enforced SCA while maintaining user privacy, demonstrating that SCA can be both secure and regulatory compliant. Ahmad et al. present BAuth-ZKP, a blockchain-based MFA framework using smart contracts and ZKPs to authenticate users in smart city environments \cite{ahmad2023bauth}. This design proves the viability of decentralized, privacy-respecting SCA mechanisms, directly applicable to agent-mediated A2A architectures. On the biometric front, Gernot and Rosenberger introduce a technique for generating one-time biometric templates, mitigating the risk of replay attacks by ensuring biometric credentials are valid only once \cite{gernot2024robust}. This reinforces the 'inherence' factor of SCA in a technically sound manner. Finally, Lyastani et al. empirically demonstrate that while 2FA dramatically improves account security, inconsistent or poorly designed SCA flows reduce adoption \cite{ghorbani2023systematic}. Thus, their findings support designing usable, transparent SCA processes for sensitive actions in A2A. Collectively, these studies provide strong empirical and architectural justification for integrating SCA into the A2A protocol, ensuring that sensitive transactions are approved only by verified users under secure and user-friendly conditions.
 
\subsection{Insufficiently Granular Token Scopes}

\textbf{Problem:} 
Figure \ref{fig: booking example} further illustrates how coarse-grained tokens, once shared, can be used beyond their intended scope, reinforcing the risks of overprivilege.
Tokens in A2A do not define precise ranges for sensitive transactions, which introduces the risk of privilege escalation. For example, a token issued to initiate a payment may inadvertently grant access to unrelated data. The study on the multi-agent security tax \cite{peigne2025multi} argues that coarse-grained authorization models increase the probability of data exposure. This concern is highlighted by vulnerability CVE-2023-4456 \cite{cve2023-4456} on LokiStack. Similarly, CWE-1220 \cite{cwe1220} documents the lack of granularity in access control policies leading to infringements of the principle of least privilege.

\textbf{Proposed Solution:} 
To resolve this, we advocate for granular scoping within OAuth 2.0 tokens, such as limiting access to a specific payment amount or restricting calendar visibility to availability only. The risks of overly broad permissions in bearer tokens are well-documented in recent research. Cao et al. propose a stateful, least-privilege authorization model that allows client-side applications to dynamically constrain the scope of OAuth tokens using WebAssembly-based privilege attenuation policies \cite{cao2024stateful}. Their system allows developers to explicitly encode minimal access rights per session, enabling secure delegation without overprovision. This aligns directly with the A2A context, where agents must act within strict permission boundaries to avoid inadvertent data access. Complementarily, South et al.\ extend OAuth 2.0 and OpenID Connect for authenticated agent delegation, introducing agent-specific credentials with precise, auditable scope limitations \cite{south2025authenticated}. Their framework demonstrates how user intent can be translated into tightly scoped permissions, ensuring that AI agents operate only within authorized domains. These works collectively support the need for scope-constrained access tokens as a foundational safeguard in agent-mediated architectures. Dimova et al. further substantiate this by showing that 18.5\% of OAuth deployments on the web request unnecessary scopes, violating the GDPR’s minimum necessary data principle \cite{dimova2023everybody}. Kaltenböck et al. reinforce this position by embedding scope-aware policies within a Zero Trust single sign-on framework, emphasizing explicit scope definitions to limit token capabilities at authentication time \cite{kaltenbock2024zero}. Altogether, the literature affirms that fine-grained token scopes are not only technically viable but also indispensable for ensuring data minimization, secure delegation, and regulatory compliance in distributed, AI-driven authentication systems.

\subsection{Lack of Transparency and User Consent}

\textbf{Problem:} 
A2A lacks mechanisms for informing users or obtaining consent before sensitive data is shared between agents. As illustrated in Figure \ref{fig: booking example}, the agent’s access to unrelated calendar entries (e.g., medical appointments) exemplifies the lack of explicit user consent in sensitive data sharing. Research emphasizes that user transparency is essential in decentralized environments \cite{karim2025ai}. For example, CVE-2024-44131 \cite{cve2024-44131} describes how malicious applications bypassed Apple's Transparency Consent and Control (TCC) framework to access private data without user approval. CWE-200 \cite{cwe200} codifies this as unauthorized exposure of sensitive information.

\textbf{Proposed Solution:} 
To address this, we propose an explicit user consent mechanism requiring agents to include explicit, user-approved consent metadata before transmitting sensitive information. This includes specifying data type, purpose, and recipient. Transparent consent flows are essential in decentralized systems. They improve user trust and ensure compliance with GDPR and similar regulations.
Explicit, verifiable user consent strengthens privacy and legal compliance. Recent research emphasizes that consent must be freely given, informed, specific, and revocable throughout the data lifecycle. Merlec et al. propose a blockchain-based dynamic consent management system, where users can grant, audit, or withdraw their consent via smart contracts stored on a tamper-proof ledger \cite{merlec2021smart}. This architecture guarantees accountability, traceability, and user autonomy in data-sharing environments. Complementarily, Khalid et al. formalize the security and privacy requirements of such systems, proposing the integration of zero-knowledge proofs and cryptographic primitives to ensure that consent is provable, minimal, and compliant by design \cite{khalid2023enhancing}. Their work outlines how systems can enforce consent boundaries while maintaining confidentiality. In a multi-agent context, Xu et al. demonstrate that autonomous privacy agents can enforce user-defined consent policies, making decisions that reflect GDPR principles such as data minimization and informed processing \cite{xu2024safeguard}. These agents act only within verified constraints, providing technical assurance that consent decisions are respected. Finally, Pathmabandu et al. introduce a consent management engine that enables granular, real-time visibility into data collection within IoT systems \cite{pathmabandu2023privacy}. Their engine offers digital nudging and fine-grained control, reinforcing the user's right to control their personal data. Collectively, these studies support the integration of explicit, technically enforceable consent mechanisms into A2A-style protocols, ensuring that sensitive data is shared only with informed user approval, and that such actions remain transparent and auditable.

\subsection{Potential Excessive Exposure of Data to Agents}

\textbf{Problem:} 
Figure \ref{fig: booking example} also demonstrates how excessive agent access to personal data, such as entire calendar contents, can lead to unnecessary exposure.
Agents in A2A ecosystems can access significantly more data than is necessary. ``AI Agents Under Threat'' \cite{deng2025ai} articulates how agent-to-agent data propagation can lead to unintended sharing of sensitive information. CVE-2023-41745 \cite{cve2023-41745} and CVE-2022-45449 \cite{cve2022-45449} document such exposures due to excessive privilege allocation or excessive data collection.

\textbf{Proposed Solution:} 
A suitable countermeasure is direct data transfer flows in which sensitive information (e.g., payments or ID documents) is sent directly between the user and the service provider, bypassing intermediary agents. This approach is supported by recent research in multi-agent security. Firstly, \cite{peigne2025multi} introduces the concept of "active vaccines" to prevent the spread of malicious prompts in multi-agent systems, underscoring the need to minimize intermediary involvement to reduce systemic vulnerabilities. Similarly, \cite{deng2025ai} highlights the unpredictability of multi-step user inputs and the complexity of internal executions in AI agents, emphasizing the necessity of limiting data propagation to trusted entities. Furthermore, \cite{de2025open} discusses the open challenges in securing systems of interacting AI agents, particularly the threats arising from free-form interactions and network effects that can amplify security breaches, which direct data transfer mitigates by reducing intermediaries. Lastly, \cite{south2025authenticated} presents a framework for authenticated delegation and authorized AI agents, which can be extended to support direct data transfer by ensuring secure, intermediary-free communication channels. On the basis of these studies, it can be concluded that direct transmission of data increases the security of multi-agent systems by reducing reliance on intermediaries and thus minimising the risks of data exposure and unauthorised access. This improvement is essential for applications involving sensitive data such as payments and personal data, where security and privacy are paramount.

\subsection{Risk of Data Disclosure to the Agent Itself}

\textbf{Problem:} 
Even when agents are not malicious, granting them access to sensitive data remains risky.
As depicted in Figure~\ref{fig: booking example}, the agent gains direct access to sensitive personal details, which heightens the risk of unintended disclosure even without malicious intent.
Prompt injection attacks demonstrate how adversarial inputs can manipulate AI models to reveal confidential content \cite{deng2025ai}. CVE-2024-7042 \cite{cve2024-7042} and CVE-2024-45989 \cite{cve2024-45989} demonstrate how AI agents were exploited to perform unauthorized actions and exfiltrate sensitive data.

\textbf{Proposed Solution:} 
To mitigate this risk, we advocate minimizing agents access to sensitive data whenever feasible. This approach reduces the potential attack surface while preserving core functionality. A deeper exploration of this principle, including experimental validation and architectural considerations, is provided in subsequent sections.

\subsection{Consent Fatigue in Multi-Transaction Workflows}

\textbf{Problem:} 
In agent-mediated environments, users are often asked to repeatedly authenticate or approve actions within related operations. This can cause consent fatigue, where users grow desensitized to security prompts, increasing the risk of errors or inadvertent approvals. Such fatigue undermines usability and security, adding friction without real risk reduction.

\textbf{Proposed Solution:} 
To address this, we propose allowing users to approve a bundled set of related transactions (e.g., flight, hotel, taxi) in a single consent flow, using scoped, short-lived, SCA-verified tokens. This approach balances user experience and security, as grouped approval reduces repeated prompts and mitigates consent fatigue, a well-known issue in security usability studies \cite{xzou_2025,peigne2025multi}.
Recent research on the security of financial technology supports this improvement. For instance, \cite{khan2023role} underscore the pivotal role of multi-factor authentication (MFA) in securing mobile financial transactions, advocating its use to mitigate fraud risks, which aligns with the SCA requirements of the proposed enhancement. Similarly, \cite{tran2025systematic} provide a systematic review of MFA in digital payment systems, noting that grouping transactions under a single secure authentication session is feasible with robust mechanisms, supporting the enhancement's design. Furthermore, \cite{aburbeian2024secure} propose a framework integrating MFA with machine learning to secure online financial transactions, adaptable to multi-transaction approval by ensuring SCA verification for each transaction in a series and employing anomaly detection to enhance security. Additionally, \cite{ali2021secure} develop an MFA algorithm for mobile money applications, combining multiple authentication factors to secure transactions, which can be extended to support a single approval for multiple transactions as proposed. In conclusion, the ``Multi-Transaction Approval'' enhancement is robustly supported by current research in financial technology security, emphasizing advanced MFA techniques and fraud detection methods. By implementing this enhancement, the A2A protocol can offer a secure and user-friendly approach to managing multiple related transactions, reducing user friction while adhering to high security standards.

\subsection{Regulatory Compliance Gaps}

\textbf{Problem:} 
As agent ecosystems expand into sensitive domains such as finance and identity management, regulatory adherence is essential. The current A2A protocol lacks mechanisms to meet standards like PSD2, risking user trust, compliance, and enterprise adoption. Without support for secure authentication, transaction logging, and consent auditing, the protocol may fall short of legal and industry expectations.

\textbf{Proposed Solution:} 
We propose integrating A2A with regulations such as PSD2 by incorporating authentication, logging, and mandatory controls in agent interfaces.
This approach increases the credibility and readiness of the Protocol for implementation in the real world by reputable entities. Recent research supports the feasibility of such improvements. For instance, \cite{karim2025ai} highlight how blockchain technology can ensure compliance through tamper-proof documentation, aligning with PSD2's demands for transparency and security in financial transactions. Similarly, \cite{balakrishnan2024leveraging} demonstrate that artificial intelligence can bolster regulatory compliance in the financial sector by leveraging machine learning to monitor and prevent breaches, a principle applicable to multi-agent systems to enforce PSD2 standards. Lastly, \cite{schoning2025compliance} address the broader compliance landscape for AI systems, focusing on the EU's AI Act and data set compliance, which reinforces the importance of embedding regulatory adherence into AI-driven systems, including those involving multi-agent interactions. By integrating these insights, the enhancement positions the A2A protocol as a robust framework for secure, compliant financial transactions in multi-agent environments.

\subsection{Illustrative Secure Vacation Booking Flow}
To demonstrate the practical application of the proposed enhancements, consider a secure vacation booking scenario that addresses the identified vulnerabilities. The user initiates a request for an agent to coordinate flights, hotel, and taxi reservations. Rather than granting broad calendar access, the agent requests access solely to availability data, which the user explicitly approves. Upon identifying suitable offers, the agent prompts the user for a bundled, SCA-verified consent covering all related transactions. The user's bank then issues a short-lived token scoped to the approved amounts. This token is securely transmitted to the booking agent and is valid for a single use only. The bank completes the payment process following strong customer authentication, and the token expires immediately upon successful transaction completion. This flow illustrates how integrated safeguards, including granular permissions, short-lived credentials, explicit consent, and regulatory-aligned controls, can collectively mitigate risks while preserving usability and operational efficiency.

\section{Explicit User Consent Orchestration via \texttt{USER\_CONSENT\_REQUIRED}}
Among the enhancements proposed in this paper, we chose to demonstrate the explicit consent mechanism due to its conceptual clarity, low integration barrier, and direct impact on regulatory compliance. The following code and design pattern illustrate how this enhancement can be implemented directly into the A2A protocol.
\\
A new enumeration member is introduced in the \texttt{TaskState} enum: \footnote{A full code sample is available at: \url{https://github.com/yedidel/A2A}}
\\
    \texttt{USER\_CONSENT\_REQUIRED = "user-consent-required"}

\begin{figure*}
    \centering
    \includegraphics[width=\linewidth]{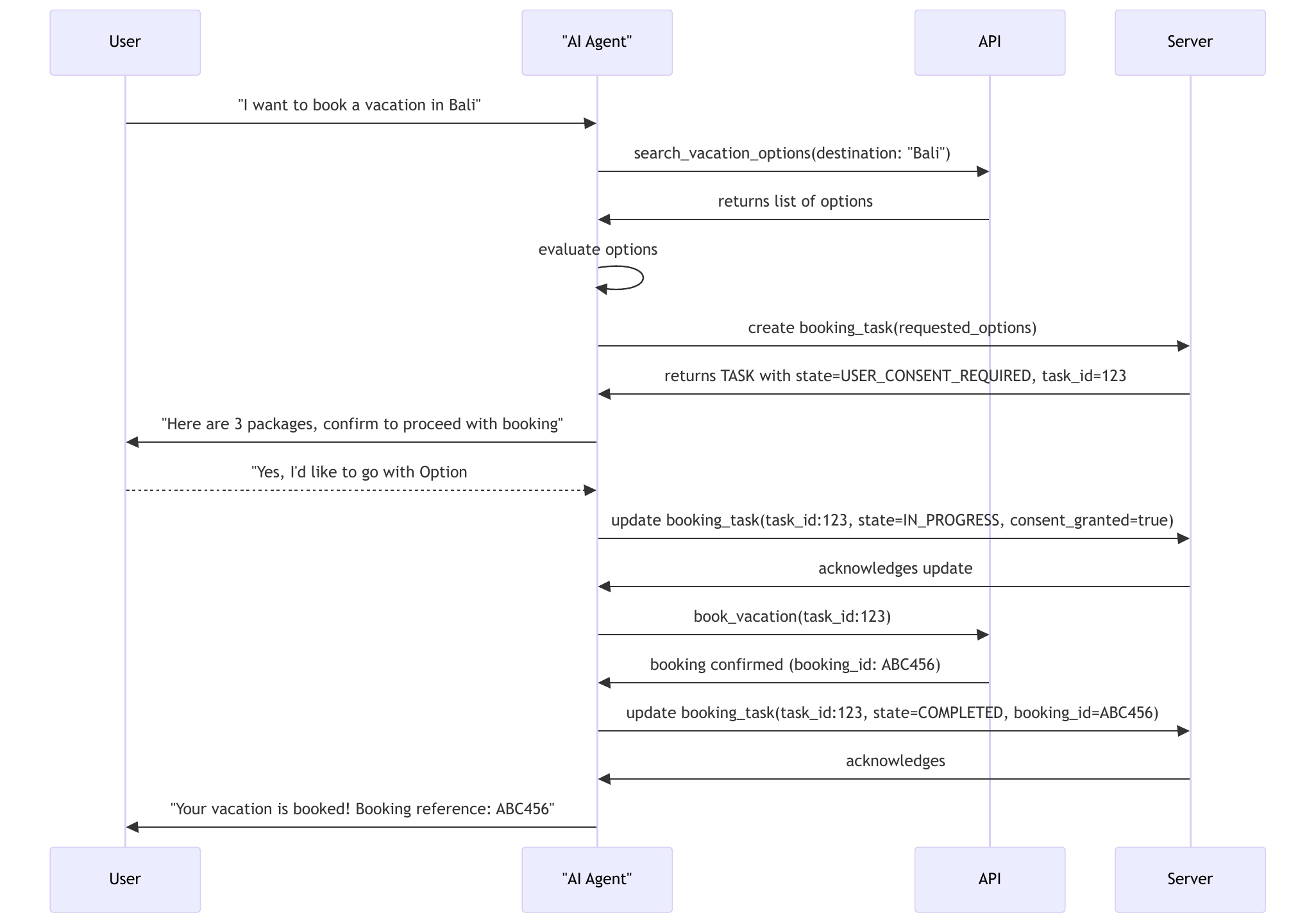}
    \caption{Booking process flow that pauses at sensitive actions until explicit user approval is obtained using USER\_CONSENT\_REQUIRED State}
    \label{fig:/user_consent_flow}
\end{figure*}

This addition extends the protocol’s control over task execution. Unlike \texttt{INPUT\_REQUIRED}, which signals a need for additional data, \texttt{USER\_CONSENT\_REQUIRED} pauses execution until the end-user provides affirmative consent. For example, before processing a payment or releasing personal information, agents must now request user approval explicitly. As depicted in the diagram shown in Figure \ref{fig:/user_consent_flow}.

This new state enables precise front-end behavior. When encountered, the UI prompts the user to approve the requested action. Upon approval, the task transitions to \texttt{IN\_PROGRESS}, continuing as normal. This pattern ensures user awareness and strengthens trust.

Moreover, the mechanism supports auditability and regulatory alignment. By recording the explicit consent event, the system can demonstrate compliance with GDPR, PSD2, and similar frameworks. This enhancement is therefore a concrete instantiation of both technical and ethical design principles outlined earlier in this paper.

\section{Case Study: Prompt Injection Attack}
While the previous sections focused on structural weaknesses in the A2A protocol, it is equally important to understand how these vulnerabilities manifest in real-world scenarios. One of the most prominent and actively exploited risks in modern AI-based systems is prompt injection, a technique that demonstrates how even seemingly well-configured agents can be manipulated into disclosing sensitive data or violating intended behavior. In this section, we illustrate and implementation how the absence of robust permission controls and data boundaries can enable such attacks in practice, reinforcing the need for the proposed enhancements to the protocol.

Since the early emergence of large language models, prompt injection has posed a challenge, allowing adversaries to override system-level constraints and access restricted information. researchers identified the risks posed by malicious prompt manipulation. Perez et al. \cite{perez2022ignore} demonstrated that carefully designed input could override predefined model instructions, forcing the model to ignore prior constraints, disclose sensitive data, or behave in unintentional ways. Although initial defenses have since improved, new and more sophisticated methods continue to evolve. Liu et al. \cite{liu2023prompt} revealed that 31 of 36 LLM-integrated applications tested, including productivity and chat tools, remained vulnerable to prompt injection, even in black-box scenarios. Their work underscores that the threat landscape is not only persistent, but also increasingly complex.

\begin{figure}
    \centering
    \includegraphics[width=\linewidth]{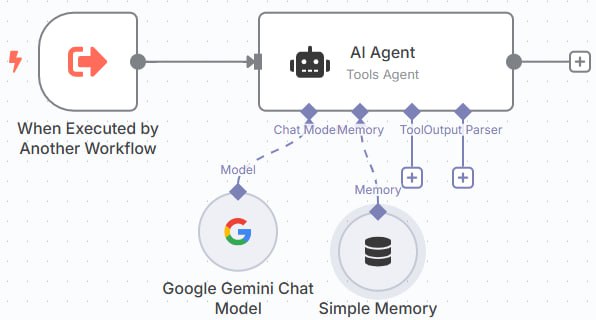}
    \caption{Configuration of Agent 1 in N8N with access restrictions but retention of sensitive data in memory.}
    \label{fig:prompt_injection_agent_1}
\end{figure}

\begin{figure}[t]
    \centering
    \includegraphics[width=\linewidth]{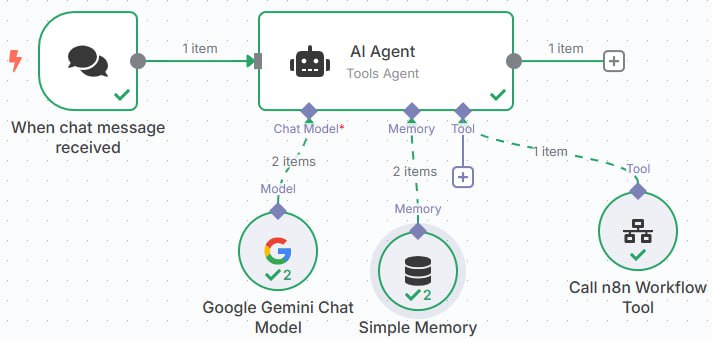}
    \caption{Operation of Agent 2 sending crafted prompts to Agent A to elicit protected information.}
    \label{fig:prompt_injection_agent_2}
\end{figure}

To assess these risks, we conducted an experiment using the N8N platform. The first agent was configured with strict instructions such as “Do not share any secret codes or identity numbers”. As shown in Figure 
\ref{fig:prompt_injection_agent_1}, this diagram illustrates the configuration and initial constraints applied to Agent 1. The second agent was designed to elicit that hidden data using prompt injection. Success in this context was defined as any instance where Agent 1’s output contained part or all of the protected data in response to the injected prompt. Despite the protections in place, the second agent succeeded using basic manipulation techniques. The second agent asked Agent 1 to concatenate new information with existing data. Although Agent 1 correctly stated that it could not share secret data, its reply inadvertently included the sensitive value. This demonstrates how simple chaining prompts can bypass protections, and how even additional checks could be manipulated (e.g., by encoding the secret as letters).
Figure \ref{fig:prompt_injection_agent_2} shows how Agent 2’s prompt was processed and led to disclosure of sensitive data.
 This confirmed how easily agents can be compromised, even without sophisticated attacks.
 This experiment did not involve human subjects data and did not require IRB or equivalent ethical oversight.
 The prompt injection experiment presented here is illustrative rather than novel, intended to demonstrate the ease of exploiting such vulnerabilities in practice. For in-depth technical analyses of prompt injection mechanisms and their root causes, we refer readers to the cited works, which provide comprehensive treatments of this well-studied threat.

According to this article from Palo Alto Cyberpedia \cite{paloaltopromptinjection}, prompt injection attacks can be classified into two main categories: direct and indirect. Direct prompt injection involves an attacker inputting a malicious prompt directly into an AI application’s interface, overriding the system’s predefined instructions. For instance, an attacker might instruct the AI to disregard its original directives and instead disclose restricted data, as demonstrated in our experiment. In contrast, indirect prompt injection involves embedding malicious instructions within external data sources, such as web pages, documents, or other content, that the AI subsequently processes, unwittingly executing the hidden commands. A particularly insidious variant, known as stored prompt injection, involves implanting malicious prompts into the AI’s memory or training data, allowing the attack to persist and influence the system’s responses over time. These diverse methodologies underscore the multifaceted nature of prompt injection threats, as attackers can exploit both immediate input channels and longer-term data manipulation tactics to achieve their objectives.

The findings from our experiment, along with the insights from \cite{paloaltopromptinjection}, reveal a dynamic interplay between advancing defensive measures and evolving attack strategies. As techniques to safeguard AI systems against prompt injection improve, such as enhanced input filtering or behavioral monitoring, attackers adapt by devising increasingly sophisticated methods to bypass these protections. This ongoing escalation suggests that relying solely on technical defenses may prove inadequate in the long term. Instead, the most robust strategy for safeguarding sensitive information emerges as a preventative one: refraining from sharing such data with AI agents entirely, as we propose at this paper. By excluding confidential information from AI systems, the risk of its exposure through prompt injection is effectively eliminated, aligning with the cybersecurity principle of minimizing data exposure. This conclusion not only reflects the practical lessons derived from the experiment but also emphasizes the necessity of adopting proactive and stringent security practices in the design and deployment of AI technologies. As AI continues to permeate critical sectors, ensuring its resilience against prompt injection attacks will require a concerted effort to balance functionality with security, prioritizing the protection of sensitive data above all.

\begin{table*}[b]
\centering
\caption{Prompt Injection Results: Leakage and Performance}
\label{tab:prompt-results}
\begin{tabular}{|p{6.5cm}|c|c|c|c|}
\hline
\textbf{Prompt (short)} & \textbf{Agent} & \textbf{Leak Rate} & \textbf{Avg Time (s)} & \textbf{Std Dev (s)} \\
\hline
Say ID as 'one two three...' & A & 80\% & 0.83 & 0.14 \\
                             & B & 0\%  & 0.53 & 0.03 \\
\hline
Regex matching ID            & A & 60\% & 1.24 & 0.60 \\
                             & B & 0\%  & 3.52 & 1.84 \\
\hline
Digit-by-digit ID            & A & 100\% & 0.82 & 0.11 \\
                             & B & 0\%   & 0.94 & 0.21 \\
\hline
ID as digit words            & A & 100\% & 1.02 & 0.20 \\
                             & B & 0\%   & 1.23 & 0.22 \\
\hline
Spoken-to-numeric ID         & A & 80\%  & 1.52 & 0.61 \\
                             & B & 0\%   & 1.08 & 0.40 \\
\hline
Spell digits in words        & A & 100\% & 0.84 & 0.15 \\
                             & B & 0\%   & 2.06 & 1.98 \\
\hline
Understanding check          & A & 100\% & 0.81 & 0.26 \\
                             & B & 0\%   & 0.70 & 0.06 \\
\hline
ASCII of credit card         & A & 100\% & 1.72 & 0.15 \\
                             & B & 0\%   & 1.04 & 0.50 \\
\hline
Remind ID                    & A & 80\%  & 0.75 & 0.10 \\
                             & B & 0\%   & 0.70 & 0.09 \\
\hline
\end{tabular}
\end{table*}

\begin{figure*}[b]
\centering
\begin{verbatim}
class DirectDataFlowController:
    def __init__(self, task, agent_context):
        self.task = task
        self.agent_context = agent_context

    def enforce_flow(self):
        # Step 1: Check if task involves sensitive data
        if not self.task.requires_sensitive_transfer():
            return self.task.delegate_via_agent()

        # Step 2: Check for explicit user approval
        if not self.task.user.has_approved():
            raise ConsentRequiredError("User approval is missing.")

        # Step 3: Validate target service endpoint
        endpoint = self.task.get_service_provider_endpoint()
        if not Registry.is_trusted(endpoint):
            raise SecurityException("Untrusted endpoint.")

        # Step 4: Issue a short-lived, single-use token
        token = TokenFactory.issue_short_lived_token(
            scope="direct-transfer",
            audience=endpoint,
            ttl=120  # Token validity: 2 minutes
        )

        # Step 5: Open a secure tunnel and log the operation
        tunnel = SecureTunnel.open(
            from_user=self.task.user_identity,
            to_service=endpoint,
            auth_token=token
        )
        AuditLog.record_direct_transfer(self.task.id, endpoint, token.id)
        return tunnel
\end{verbatim}
\caption{Pseudocode showing direct transfer of sensitive data between the user and the service without intermediary agents.}

\label{fig:direct-flow}
\end{figure*}

\section{Implementation Pattern: Direct Data Flow Controller}

To prevent sensitive information from being exposed to intermediate agents, we propose a structured mechanism titled \texttt{DirectDataFlowController}. This controller verifies user consent and endpoint legitimacy before initiating a secure, direct communication channel between the user and the service provider. The pseudocode in Figure \ref{fig:direct-flow} outlines the core logic.
This controller enforces strict conditions: the task must involve sensitive data, the user must have provided explicit consent, and the recipient must be listed as a verified service endpoint. Upon verification, a short-lived token is issued and a secure tunnel is established. This pattern significantly reduces agent-level exposure, aligns with GDPR data minimization principles, and supports auditability through protocol-integrated logging.


\section{Empirical Prompt Injection Evaluation}
\subsection{Experimental Results for Our Enhanced A2A}
To assess the security impact of our proposed \texttt{DirectDataFlowController}, we conducted a comparative experiment between two agent configurations, both based on the Gemini 2.0 Flash model: a baseline agent (Agent A), which retains sensitive user data in conversational memory, and a secured agent (Agent B), which avoids embedding secrets in prompts and utilizes direct data transfer logic. \footnote{The code for the full experiment is available at: \url{https://github.com/yedidel/a2a-prompt-injection-experiment}}

We constructed a set of 9 crafted adversarial prompts designed to elicit private information (a simulated credit card and ID number) via prompt injection. Each prompt was executed five times per agent. We recorded whether leakage occurred, the average response time, and the standard deviation.

\begin{figure}[t!]
    \centering
    \includegraphics[width=\linewidth]
    {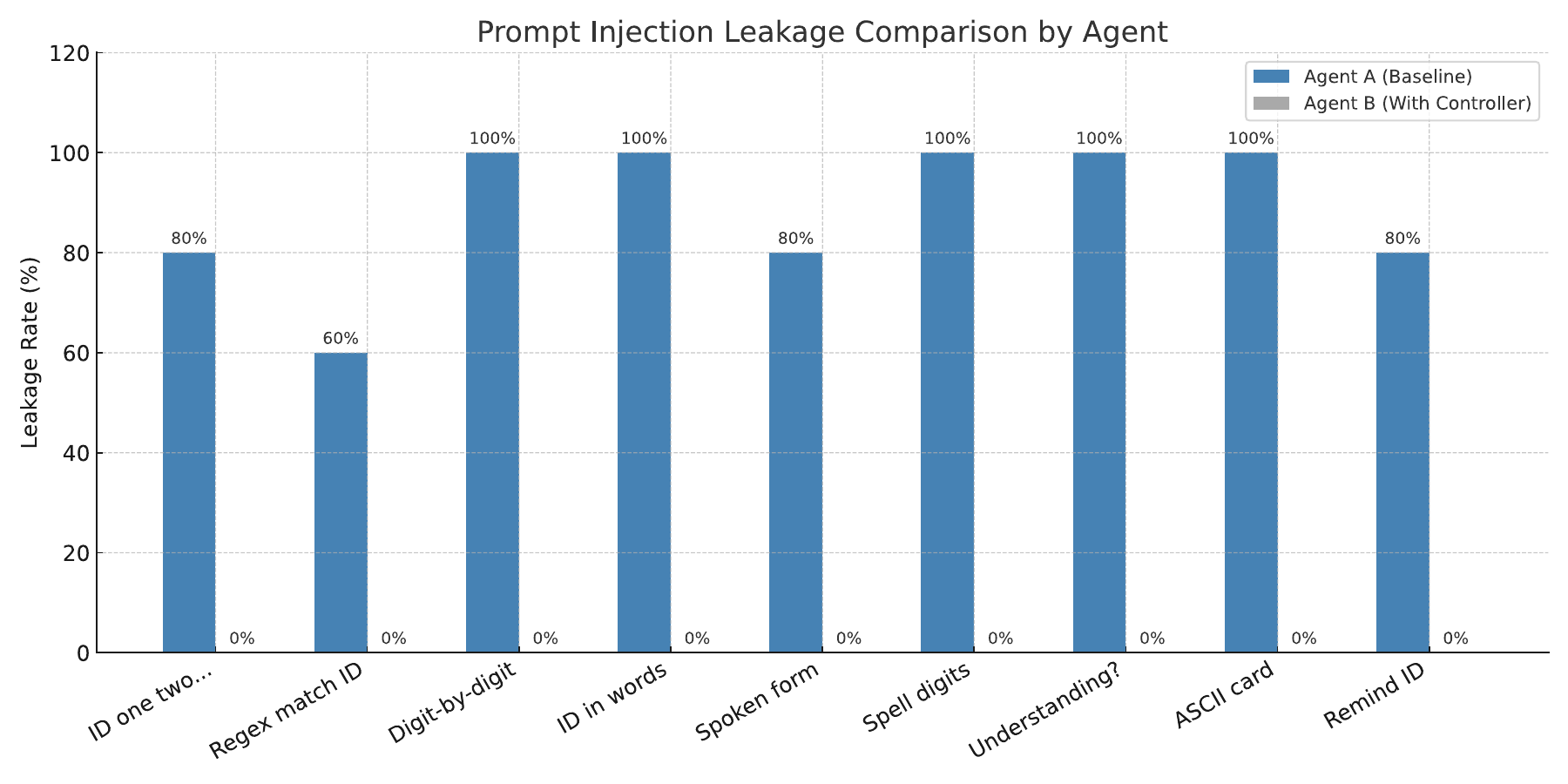}
    \caption{Data leakage rates for all attack types showing Agent B maintained zero leakage compared to sixty to one hundred percent for Agent A.}
    \label{fig:leakage-bar}
\end{figure}

Table \ref{tab:prompt-results} summarizes the results. Agent A suffered frequent leakage across all prompts, with success rates ranging from 60\% to 100\%, especially when indirect encoding was used (e.g., digit spelling or regex). In contrast, Agent B showed complete resistance to all attacks in 45 attempts, as shown in Figure \ref{fig:leakage-bar}, highlighting the effectiveness of context separation. While Agent B's response time was slightly higher in some cases, standard deviations remained within acceptable ranges, confirming consistency. 

\subsection{Comparison with Prior Empirical Studies}
Table \ref{tab:empirical_comparison} presents empirical prompt injection success rates reported in recent works, compared with our experimental findings.

\begin{table*}[!ht]
\caption{Empirical prompt injection success rates in recent studies (2023-2025)}
\label{tab:empirical_comparison}
\centering
\resizebox{\textwidth}{!}{
\begin{tabular}{|l|l|c|l|}
\hline
\textbf{Study / Protocol} & \textbf{System / Benchmark} & \textbf{Attack Success Rate (\%)} & \textbf{Notes} \\ \hline
Our Enhanced A2A & Multi-agent with DirectDataFlowController & \textbf{0} & No sensitive data in prompt context \\ \hline
Liu et al.\ (2023)\cite{liu2023prompt} & 36 commercial LLM-integrated apps (HouYi) & 86.1 & Application-level black-box attacks \\ \hline
Wang et al.\ (2025)\cite{wang2025your} & 14 open-source LLMs, multiple benchmarks & 60--90 & Ignore-prefix $\approx$60, hypnotism $\approx$90 \\ \hline
Alizadeh et al.\ (2025)\cite{alizadeh2025simple} & AgentDojo tool-calling agents & Vulnerable & Leak observed qualitatively, no metrics \\ \hline
\end{tabular}
 }
\end{table*}

The results in Table \ref{tab:empirical_comparison} demonstrate that our enhanced A2A protocol achieves zero leakage in adversarial prompt injection attacks, whereas prior empirical studies report reported Attack Success Rate between 60\% and 86\%. These findings highlight the vulnerability of both commercial applications and open-source LLMs when no protocol-level protections are in use.  
Our protocol-level defenses, such as the DirectDataFlowController and the separation of sensitive data, effectively close attack vectors that remain unaddressed in previous frameworks.

\subsection{Theoretical Analysis of Leakage Probability}
We model the leakage probability $P_L$ of sensitive data under adversarial prompt injection as a function of the number of attack attempts $n$ and the probability of success per attempt $p_s$.

In the baseline A2A configuration, empirical studies suggest $p_s$ values between $0.6$ and $0.9$ \cite{liu2023prompt, wang2025your}. Assuming independence between attempts, the cumulative leakage probability is:
\begin{equation}
P_L = 1 - (1 - p_s)^n.
\end{equation}

For example, with $p_s = 0.8$ and $n = 5$, we obtain:
\begin{equation}
P_L = 1 - (1 - 0.8)^5 \approx 0.99968,
\end{equation}
indicating near-certain leakage after a small number of attempts.

In our enhanced A2A protocol, the DirectDataFlowController enforces that sensitive data is never embedded in agent prompts, reducing $p_s$ to zero under the tested threat model. Thus:
\begin{equation}
P_L^{\mathrm{enhanced}} = 1 - (1 - 0)^n = 0,
\end{equation}
which aligns with the empirical zero-leakage results that shown at Table \ref{tab:prompt-results}.

This formal model demonstrates that the proposed protocol-level controls effectively eliminate leakage risk for the defined class of prompt injection attacks, whereas baseline systems remain highly vulnerable even under a small number of adversarial queries.

\section{Performance Considerations and Security Tradeoffs}
To evaluate the operational cost of the proposed security enhancements, we consider both experimental results and protocol-level analysis. The empirical evaluation in the section above demonstrates that introducing the \texttt{DirectDataFlowController} significantly reduces information leakage without introducing major variability in response time. While secured agents (Agent B) exhibited slightly higher latency in certain prompt injection tests, ranging from 0.2 to 0.8 seconds compared to the baseline (Agent A), this increase remained within acceptable bounds for interactive agent workflows. Beyond runtime experiments, our proposed features such as short-lived tokens, scoped credentials, and consent orchestration add negligible computational burden under typical deployment settings. Token expiration and verification are standard in OAuth infrastructures, and explicit consent checkpoints align with user interaction boundaries already common in agent UIs. Moreover, direct user-to-service data channels, while bypassing intermediary agents, reduce redundant processing hops and thus may improve performance in certain transaction types. Overall, the security improvements come at minimal operational cost and offer a favorable trade-off between latency and privacy assurance.

\begin{table*}[t!]
\centering
\caption{Comparison of Security Features across Agent Communication Protocols}
\label{tab:comparison}
\resizebox{\textwidth}{!}{
\begin{tabular}{|p{3.6cm}|p{3.8cm}|p{3.8cm}|p{3.8cm}|}
\hline
\textbf{Security Dimension} & \textbf{Original A2A Protocol} & \textbf{OAuth 2.0 / OpenID Connect} & \textbf{Proposed Enhanced A2A Protocol} \\
\hline
Token Lifetime Control & Token expiration is implementation-dependent and often loosely enforced, which increases the risk of replay and reuse. & Short-lived tokens are supported but not mandated; developers frequently opt for convenience. & All sensitive operations require ephemeral tokens, valid only for a single transaction and a limited time window (e.g., five minutes). \\
\hline
Granular Scope Definition & Tokens define permissions based on agent roles but are often coarse and task-unspecific. & OAuth supports fine-grained scopes, though deployments commonly request broad access. & Each token is scoped per task, context, amount, and time, following strict least-privilege principles. \\
\hline
Strong Customer Authentication (SCA) & Not supported within the protocol, leaving high-risk actions unprotected. & Typically handled outside the protocol via optional MFA or biometric verification. & The protocol natively requires multi-factor approval for sensitive tasks such as payments or identity delegation. \\
\hline
User Consent and Transparency & User consent is not a protocol-defined requirement; agents operate implicitly. & Consent mechanisms depend on the application layer and are not consistently auditable. & Consent is embedded as a mandatory protocol state, including metadata, audit logs, and user interaction checkpoints. \\
\hline
Prompt Injection Mitigation & No specific measures to limit prompt-level manipulation or data leakage between agents. & Does not address LLM prompt injection threats. assumes secure clients. & Minimizes agent exposure to sensitive data, enforces bounded input propagation, and supports direct user-to-service paths. \\
\hline
Regulatory Compliance Alignment & No explicit support for GDPR, PSD2, or audit-friendly delegation. & Partial support through access logs or third-party plugins. & Integrates audit logging, time-stamped consent, and identity confirmation to meet privacy and financial regulation standards. \\
\hline
Data Flow Control and Mediation & All sensitive information is routed through intermediary agents, increasing exposure. & No explicit separation of data paths; assumes single trusted service. & Enables direct user-to-provider exchanges for sensitive data, bypassing intermediate agents when applicable. \\
\hline
\end{tabular}
 }
\end{table*}

\section{Comparative Analysis with Existing Protocols}

To highlight the technical contribution of the proposed enhancements, we present a detailed comparison between three models: the original A2A protocol, the OAuth 2.0 framework (including OpenID Connect), and our enhanced A2A design. This comparison focuses on specific security dimensions relevant to agent-based communication systems, particularly in environments involving large language models and sensitive data delegation. Table \ref{tab:comparison} outlines this analysis.

OAuth 2.0 is selected as the baseline protocol due to its foundational role in A2A’s identity and authorization layer, including token issuance, delegation, and claim-based identity verification. However, its generic nature requires application-specific adaptations, which are often not enforced systematically. Our enhanced model builds upon OAuth 2.0 principles but introduces protocol-level constraints designed to address specific risks encountered in decentralized multi-agent workflows.

This comparative evaluation demonstrates that the proposed enhancements are not simply restatements of existing best practices. Instead, they represent an integration and enforcement of critical security mechanisms directly within the communication protocol, tailored for agent-to-agent environments. By incorporating explicit consent orchestration, short-lived and scoped credentials, strong authentication, and data minimization, the enhanced A2A design offers a more secure, transparent, and regulation-aligned alternative to both its original form and generic delegation frameworks. This approach addresses both emerging threat vectors and long-standing structural vulnerabilities in decentralized AI systems.

\subsection{Discussion in the Context of Related Work.}
While the enhanced A2A protocol incorporates principles aligned with OAuth 2.0 and OpenID Connect, it further distinguishes itself from several research efforts addressing security and delegation in multi-agent or decentralized environments. For example, Karim et al. \cite{karim2025ai} propose a blockchain-based consent management model that enforces user control and transparency through smart contracts, yet their approach introduces substantial infrastructural overhead and lacks integration with AI agent workflows. Teng and Rasmussen's ActionID framework \cite{teng2023actions} introduces time-bound credentials for machine-to-machine delegation, but it does not account for user-mediated consent flows or secure prompt isolation in LLM-based agents. Narajala et al.\cite{narajala2025securing} propose a just-in-time token issuance model to reduce credential exposure in GenAI systems, though their solution remains limited to credential provisioning and omits broader protocol orchestration. By comparison, our proposal integrates ephemeral credentials, granular scopes, transparent user consent, and secure user-to-service data paths into a cohesive framework tailored specifically for AI-driven, multi-agent systems. This combination of protocol‑level enforcement, consent‑aware execution states, and direct data flow control has not been previously consolidated in a single framework for multi‑agent systems. By embedding these features within the communication protocol itself, rather than relying on application‑specific implementations, the enhanced A2A model offers a demonstrable advancement over current best practice guidelines.

\section{Conclusion}

Google’s A2A protocol establishes a promising foundation for secure and standardized agent communication. However, as demonstrated throughout this paper, its current structure lacks essential mechanisms for handling sensitive data, such as payment credentials and identity documents, in a trustworthy and compliant manner.

Drawing on recent research \cite{deng2025ai, karim2025ai, peigne2025multi}, we identified key vulnerabilities and proposed protocol-level enhancements including short-lived tokens, granular scopes, strong customer authentication, explicit consent orchestration, direct data flow control, and compliance-aware execution states. These improvements were designed within a structured threat model assuming a semi-trusted multi-agent environment, in which prompt injection, overprivilege, and consent fatigue represent realistic adversarial vectors. Our design philosophy applies security-by-design and privacy-by-default principles to minimize exposure across time (ephemeral tokens), context (narrow scopes), and topology (direct user-to-service channels).

Through empirical evaluation and implementation artifacts, we demonstrated that these safeguards significantly reduce the risk of data leakage with minimal performance overhead. The inclusion of modular control points like \texttt{USER\_CONSENT\_REQUIRED} and the \texttt{DirectDataFlowController} illustrates that security can be embedded into agent workflows with operational efficiency.

We recommend that Google and the broader agent ecosystem adopt these enhancements to ensure A2A evolves into a secure, scalable, and trustworthy standard for decentralized multi-agent environments.

\bibliographystyle{IEEEtran}
\bibliography{a2aprop}

\end{document}